# THERMOELASTIC RESPONSE OF BRAGG CRYSTALS UNDER MHz THERMAL LOADING*

P. Liu, K.-J. Kim, R. Lindberg, Yu. Shvyd'ko, Argonne National Laboratory, Lemont, USA

## Abstract

An x-ray free-electron laser oscillator (XFELO) is a promising candidate for producing fully coherent x-rays beyond the fourth-generation light sources. An R&D XFELO experiment [1–5] (ANL-SLAC-Spring-8 collaboration) to demonstrate the basic principles and measure the two-pass FEL gain is expected to be accomplished by 2025. Beyond this R&D experiment, an XFELO user facility will be eventually needed to produce stable x-ray pulses with saturated pulse energy at MHz repetition rate. One of the outstanding issues for realizing an MHz XFELO is the possible Bragg crystal degradation due to the high-repetition-rate thermal loading from high-pulse-energy x-rays. The deposited energy by one x-ray pulse induces temperature gradients and elastic waves in the crystal, where the deformed crystal lattice impacts the Bragg performance for subsequent x-ray pulses. Here, we present studies of the crystal thermoelastic response under thermal loading of high-energy x-ray pulse trains.

## INTRODUCTION

The success of x-ray free-electron lasers (XFELs) and fourth-generation light sources has paved the way for diverse science applications [6–8]. On the other hand, the thermal loading of high-energy and high-repetition-rate x-ray pulses brings challenges to stable operations of x-ray optical components such as monochromators and crystals. In particular, for future cavity-based XFELs such as an XFELO [1–3, 9], the mJ-energy x-ray pulses could introduce significant optics angular or energy shifts due to crystal thermoelastic responses.

For elastic crystals under thermal stresses, the crystal displacement field $u(t, x, y, z) = (u_x, u_y, u_z)$ and temperature field $T(t, x, y, z)$ can be described by the classical coupled thermoelastic equations [10]:

$$k\nabla^2 T = \rho c_V \dot{T} + \alpha T_0 (3\lambda + 2\mu)\nabla \cdot \dot{u} - R, \quad (1)$$

$$\mu\nabla^2 u + (\lambda + \mu)\nabla\nabla \cdot u - (3\lambda + 2\mu)\alpha\nabla(T - T_0) = \rho\ddot{u}. \quad (2)$$

Here, Lamé parameters $\lambda$ and $\mu$ are related to Poisson's ratio $\nu = \lambda/2(\lambda+\mu)$ and Young's modulus $Y = \mu(3\lambda+2\mu)/(\lambda+\mu)$. $k, \rho, c_V$, and $\alpha$ are the thermal conductivity, density, specific heat, and coefficient of linear expansion, respectively.

The mechanical coupling $\nabla \cdot \dot{u}$ term in Eq. (1) typically has a very small contribution to the temperature field, which usually leads to the dissipation of the elastic wave amplitude.

In particular, for diamond crystals in an XFELO, this term is a few orders weaker than other terms and can be neglected when not considering thermoelastic damping.

## TEMPERATURE FIELD

Without the mechanical coupling term, the thermal and elastic equations are decoupled. We can first solve the crystal temperature field independently, and then apply the temperature solution to find out the crystal displacement. Without internal heat source ($R = 0$), Eq. (1) turns to the classical heat equation:

$$\rho c_V \frac{\partial T}{\partial t} = k\nabla^2 T. \quad (3)$$

The Bragg reflection of one x-ray pulse and the subsequent heat deposition into the crystal usually have a very short time constant (typically ps). On the other hand, the thermal diffusion usually takes much longer time, typically μs or longer. Compared to this time scale, the x-ray interaction and local heat deposition can be considered instantaneous. The deposited local thermal energy density $I(x, y, z)$ can be described as

$$I(x, y, z) = \frac{Q_T}{2\pi\sigma_x\sigma_y\zeta(1 - e^{-\frac{d}{\zeta}})} e^{-\frac{(x-\mu_x)^2}{2\sigma_x^2} - \frac{(y-\mu_y)^2}{2\sigma_y^2} - \frac{z}{\zeta}}, \quad (4)$$

where $\mu_x(\mu_y)$ is the position of the x-ray beam center, $\sigma_x(\sigma_y)$ is the beam spot size in $x(y)$-direction, and $d$ is the crystal thickness. $\zeta$ is the crystal absorption length (if absorption is dominant) or extinction length (if x-ray reflection is dominant). $Q_T$ is the total deposited thermal energy. In extinction regime, $Q_T = (1 - R)Q$ with $Q$ being the incoming x-ray pulse energy and $R$ being the x-ray reflectivity. In absorption regime, $Q_T \approx Q(1 - e^{-d/\zeta})$.

The heat equation Eq. (3) with an initial temperature distribution can be solved analytically. For example, the temperature inside a 2D infinite plane with constant thermal properties and a 2D Gaussian incident beam profile has been considered previously [11, 12]. This can be extended to a 3D infinite thick plate with a 3D Gaussian thermal energy density deposition $I(x, y, z) = Q_T \mathcal{N}(0, \sigma_x^2)\mathcal{N}(0, \sigma_y^2)\mathcal{N}(0, \zeta^2)$ with $\mathcal{N}(\mu, \sigma^2) = \frac{e^{-(x-\mu)^2/2\sigma^2}}{\sqrt{2\pi}\sigma}$. The semi-infinite diamond is mirrored at $z = 0$ plane to ensure no net heat transfer at boundary $z = 0$. Assuming the crystal is initially at steady temperature $T_0$ and denoting $T$ as the temperature relative to $T_0$, the initial condition is

$$T(0, x, y, z) = \frac{Q_T}{\rho c_V}\mathcal{N}(0, \sigma_x^2)\mathcal{N}(0, \sigma_y^2)\mathcal{N}(0, \zeta^2), \quad (5)$$

* WORK AT ANL IS SUPPORTED BY ANL LABORATORY DIRECTED RESEARCH AND DEVELOPMENT (LDRD) PROJECT PRJ1010218 AND THE U.S. DEPARTMENT OF ENERGY, OFFICE OF SCIENCE, OFFICE OF BASIC ENERGY SCIENCES, UNDER CONTRACT NO. DE-AC02-06CH11357.

With boundary condition $T|_\infty = 0$ and no heat exchange at boundaries, the temperature field after one pulse is given by

$$T(t, x, y, z) = T(0, x, y, z) * \frac{e^{-\frac{x^2+y^2+z^2}{4Dt}}}{(4\pi Dt)^{\frac{3}{2}}}, \quad (6)$$

which is still a Gaussian distribution. Here, $D = k/\rho c_V$ is the thermal diffusivity, and "*" refers to convolution.

By changing the initial temperature in Eqs. (5) and (6), the temperature after the $n$ pulses (with pulse separation $\tau$) is given by

$$T = \frac{1}{\rho c_V} \sum_{m=0}^{n-1} Q_{T,m} \mathcal{N}\left(0, \sigma_x^2 + 2Dt'\right) \mathcal{N}\left(0, \sigma_y^2 + 2Dt'\right)$$
$$\times \mathcal{N}\left(0, \zeta^2 + 2Dt'\right) \Theta(t'), \quad t' = t - m\tau. \quad (7)$$

Here, $\Theta(t) = 1$ for $t > 0$ and $\Theta(t) = 0$ otherwise.

In particular, assuming x-rays have the same pulse energy $Q$, the temperature at the beam center $\mathbf{0} = (0, 0, 0)$ right before $n + 1$ pulse comes, defined as residue temperature, is given by,

$$T(n\tau, \mathbf{0}) = \frac{Q_T}{\sqrt{(2\pi)^3} \rho c_V} \sum_{m=1}^{n} \frac{1}{\sqrt{\sigma_x^2 + 2Dm\tau}} \frac{1}{\sqrt{\sigma_y^2 + 2Dm\tau}}$$
$$\times \frac{1}{\sqrt{\zeta^2 + 2Dm\tau}} < \frac{Q_T}{\sqrt{(4\pi D\tau)^3} \rho c_V} \sum_{m=1}^{n} \frac{1}{m^{3/2}}. \quad (8)$$

Clearly, the residue temperature converges as $n$ goes to infinity, which means after long enough time, x-ray pulses will interact with the crystal with the same thermal properties. Since Riemann zeta function $\zeta(1.5) = \sum_{m=1}^{\infty} \frac{1}{m^{3/2}} \approx 2.6$, the stabilized residue temperature $T_s$

$$T_s(n\tau, \mathbf{0}) < \frac{3Q_T}{\sqrt{(4\pi D\tau)^3} \rho c_V}. \quad (9)$$

Note that in the 2D case, this residue temperature (proportional to $\sum_{m=1}^{n} \frac{1}{m}$) does not converge.

For the thermal density in Eq. (4), $I(x, y, z) = 2Q_T \mathcal{N}(0, \sigma_x^2) \mathcal{N}(0, \sigma_y^2) \text{Laplace}(0, \zeta)$, i.e., normally distributed in $x$- and $y$-directions, but Laplace distributed in $z$-direction. The convolution of a Gaussian distribution $\mathcal{N}(\mu, \sigma^2)$ with a Laplacian distribution $\text{Laplace}(0, \zeta)$, denoted as $\mathcal{N}_L(\mu, \sigma^2, \zeta)$, is

$$\mathcal{N}_L(\mu, \sigma^2, \zeta) = \frac{1}{2}\left[\text{EMG}\left(z; \mu, \sigma^2, \frac{1}{\zeta}\right) + \text{EMG}\left(-z; \mu, \sigma^2, \frac{1}{\zeta}\right)\right], \quad (10)$$

where the exponentially modified Gaussian distribution

$$\text{EMG}(z; \mu, \sigma^2, \lambda) = \frac{\lambda}{2} e^{\frac{1}{2}\lambda^2 \sigma^2} e^{-\lambda(z-\mu)} \text{Erfc}\left(\frac{\lambda\sigma^2 - z + \mu}{\sqrt{2}\sigma}\right). \quad (11)$$

Using this convolution, the crystal temperature after $n$ pulses can be written as,

$$T(t, x, y, z) = \frac{2Q_T}{\rho c_V} \sum_{m=0}^{n-1} \mathcal{N}\left(0, \sigma_x^2 + 2Dt'\right) \mathcal{N}\left(0, \sigma_y^2 + 2Dt'\right)$$
$$\times \mathcal{N}_L\left(0, 2Dt', \zeta\right) \Theta(t'), \quad t' = t - m\tau. \quad (12)$$

As the scaled complementary error function $e^{x^2}\text{Erfc}(x) < 1/\sqrt{\pi}x$, the residue temperature

$$T(n\tau, \mathbf{0}) = \frac{2Q_T}{4\pi\zeta\rho c_V} \sum_{m=1}^{n} \frac{e^{\frac{Dm\tau}{\zeta^2}}\text{Erfc}\left(\frac{\sqrt{Dm\tau}}{\zeta}\right)}{\sqrt{\sigma_x^2 + 2Dm\tau}\sqrt{\sigma_y^2 + 2Dm\tau}} \quad (13)$$

$$< \frac{2Q_T}{\sqrt{(4\pi D\tau)^3} \rho c_V} \sum_{m=1}^{n} \frac{1}{m^{3/2}}. \quad (14)$$

is also a converging series.

Figure 1 shows a simulation of the temperature inside a diamond crystal plate with 100 x-ray pulses and a constant temperature cooling at one edge. As can be seen, after 100 pulses, the residue temperature (baseline) is close to be stabilized.

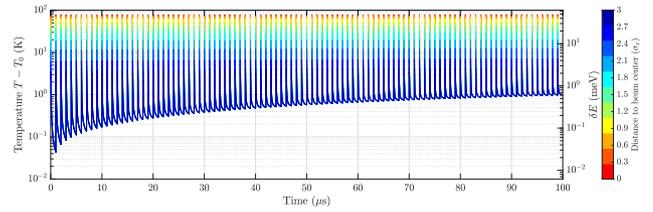

Figure 1: Simulation of crystal temperature (at different transverse locations) after 100 x-ray pulses in a diamond crystal plate with $T_0 = 100$ K, $Q = 0.5$ mJ, $R = 0.99$, $\sigma_x = \sigma_y = 44$ μm, $\zeta = 20$ μm, $\tau = 1$ μs.

## STRAIN FIELD

The strain field can be solved using equation Eq. (2) given the temperature field. To characterize shifts of Bragg reflection due to atomic plane motion in the longitudinal $z$-direction, we consider a simplified 1D longitudinal case with negligible Poisson ratio (for diamond crystal, Poisson ratio $\nu \ll 1$). Then Eq. (2) becomes

$$\frac{1}{c_L^2}\frac{\partial^2 u}{\partial t^2} - \frac{\partial^2 u}{\partial z^2} = -\alpha \frac{\partial}{\partial z}\Delta T(z, t). \quad (15)$$

Here $c_L = \sqrt{Y/\rho}$ is the speed of wave propagation. The strain field ($\epsilon_z = \partial u/\partial z$), without considering the inertial term ($\partial^2 u/\partial t^2$), is $\epsilon_z = \alpha\Delta T(z, t)$, which means the strain field has the same damping as the temperature field. The Bragg energy shift $|\delta E/E|$ is approximately proportional to $\epsilon_z$, i.e., $|\delta E/E| = \alpha\Delta T$. This is shown as the second $y$-axis in Fig. 1, with $\alpha = 4.3 \times 10^{-8}$ [13] for diamond crystal at 100 K and $E = 14.41$ keV for C(733) reflection.

For transient strain analysis, the inertial term cannot be neglected. Eq. (15) has to be considered with proper boundary (stress-free) and strain-free initial conditions. A strain wave solution can be analytically found similar to [14] but with a time-dependent temperature field: $\epsilon_z = \alpha \Delta T(z,t) + F(z,t)$, with $F(z,t)$ term characterizing the wave propagation. Figure 2 shows the strain wave propagation inside an infinite thick diamond crystal at different time instants.

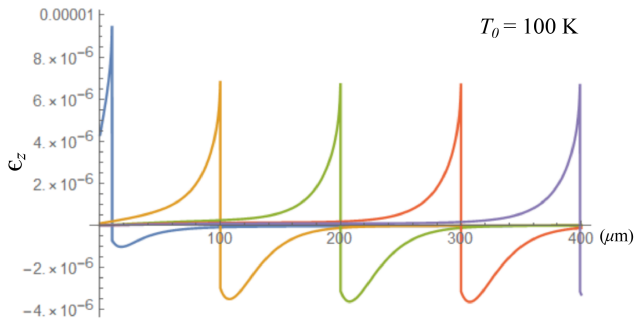

Figure 2: The propagation of strain wave $\epsilon_z$, excited by a 0.5-mJ x-ray pulse in an infinite thick diamond crystal. Curves of different colors represent $\epsilon_z$ at different time instants $c_L t = 10, 100, 200, 300, 399\ \mu m$.

## DISCUSSION

The crystal's thermoelastic shift could significantly impact the FEL buildup process in an XFELO. The diamond Bragg reflection curve can be distorted and shifted, leading to a modulation of the cavity gain or loss and spectral bandwidth. The larger the thermal loading from one pulse, the larger modulation it brings to the cavity, which in turn tends to reduce the thermal loading by the next pulse due to reduced gain or increased loss. Such a natural feedback mechanism in this dynamic process between the thermal loading and the FEL buildup may lead to stable, periodic, or even chaotic behavior of the FEL pulses, similar to other types of FEL oscillators [15–17].

The combination of crystal thermal loading with the FEL simulation has been investigated in previous studies [18–21]. It was found that instead of a stable cw structure in saturation (without considering the thermal loading), the FEL may have an oscillating or period macroscopic pulsing structure. While cryo-generically cooling of the diamond crystal can provide improved thermal damping, an efficient mechanical damping and a fast feedback mechanism are yet to be studied to suppress the unstable pulse structures of an XFELO.

## ACKNOWLEDGMENTS

We would like to thank Lahsen Assoufid and Marion White (ANL) for helpful discussions.